\documentclass[times, review, 12pt]{elsarticle}

\usepackage{booktabs} 
\usepackage[ruled]{algorithm2e} 
\usepackage{graphicx}
\usepackage{color, soul}  
\usepackage{enumitem}
\usepackage[normalem]{ulem}
\usepackage{caption}
\usepackage[caption=false,font=footnotesize,labelformat=simple]{subfig}
\usepackage{amsmath}
\usepackage{framed}
\usepackage[table]{xcolor}
\definecolor{lightgray}{gray}{0.9}

\usepackage{pdfpages}
\usepackage{supertabular}

\usepackage{amssymb}

\journal{Information and Software Technology}

\begin{document}

\begin{frontmatter}




\title{\textit{Where Are The Gaps?} A Systematic Mapping Study of Infrastructure as Code Research}


\author{Akond Rahman, Rezvan Mahdavi-Hezaveh, and Laurie Williams}

\address{North Carolina State University, Raleigh, NC, USA}

\begin{abstract}
\textit{Context:} Infrastructure as code (IaC) is the practice to automatically configure system dependencies and to provision local and remote instances. Practitioners consider IaC as a fundamental pillar to implement DevOps practices, which helps them to rapidly deliver software and services to end-users. Information technology (IT) organizations, such as Github, Mozilla, Facebook, Google and Netflix have adopted IaC. A systematic mapping study on existing IaC research can help researchers to identify potential research areas related to IaC, for example, the areas of defects and security flaws that may occur in IaC scripts.   \\
\textit{Objective: The objective of this paper is to help researchers identify research areas related to infrastructure as code (IaC) by conducting a systematic mapping study of IaC-related research.} \\
\textit{Methodology:} We conduct our research study by searching six scholar databases. We collect a set of 33,887 publications by using seven search strings. By systematically applying inclusion and exclusion criteria, we identify 31 publications related to IaC. We identify topics addressed in these publications by applying qualitative analysis. \\
\textit{Results:} We identify four topics studied in IaC-related publications: (i) framework/tool for infrastructure as code; (ii) use of infrastructure as code; (iii) empirical study related to infrastructure as code; and (iv) testing in infrastructure as code. According to our analysis, 52\% of the studied 31 publications propose a framework or tool to implement the practice of IaC or extend the functionality of an existing IaC tool. \\
\textit{Conclusion:} Our findings suggest that framework or tools is a well-studied topic in IaC research. As defects and security flaws can have serious consequences for the deployment and development environments in DevOps, along with other topics, we observe the need for research studies that will study defects and security flaws for IaC.  
\end{abstract}



\end{frontmatter}




\section{Introduction}
\label{intro}

Infrastructure as code (IaC) is the practice to automatically configure system dependencies and to provision local and remote instances~\cite{Humble:2010:CD}. Use of IaC scripts is essential to the implementation of the practice of automated deployment, such as is done with a continuous deployment process. Popular IaC technologies, such as Chef~\footnote{https://www.chef.io/chef/} and Puppet~\footnote{https://puppet.com/}, provide utilities to automatically configure and provision software deployment infrastructure using cloud instances. Information technology (IT) organizations such as, Ambit Energy~\cite{ambit:pup}, Github~\footnote{https://speakerdeck.com/kpaulisse/puppetconf-2016-scaling-puppet-and-puppet-culture-at-github}, Mozilla~\cite{cd:adage:parnin}, and Netflix~\cite{cd:adage:parnin} use these utilities to provision cloud-based instances, such as Amazon Web Services (AWS)~\footnote{https://aws.amazon.com/}, managing databases, and managing user accounts both on local and remote computing instances. For example, Puppet provides the `sshkey resource' to install and manage secure shell (SSH) host keys and the `service resource' to manage software services automatically~\cite{puppet-doc}. Use of IaC scripts has helped IT organizations to increase their deployment frequency. For example, Ambit Energy, uses IaC scripts to increased their deployment frequency by a factor of 1,200~\cite{ambit:pup}. 

\begin{figure}
\includegraphics[width=0.9\textwidth]{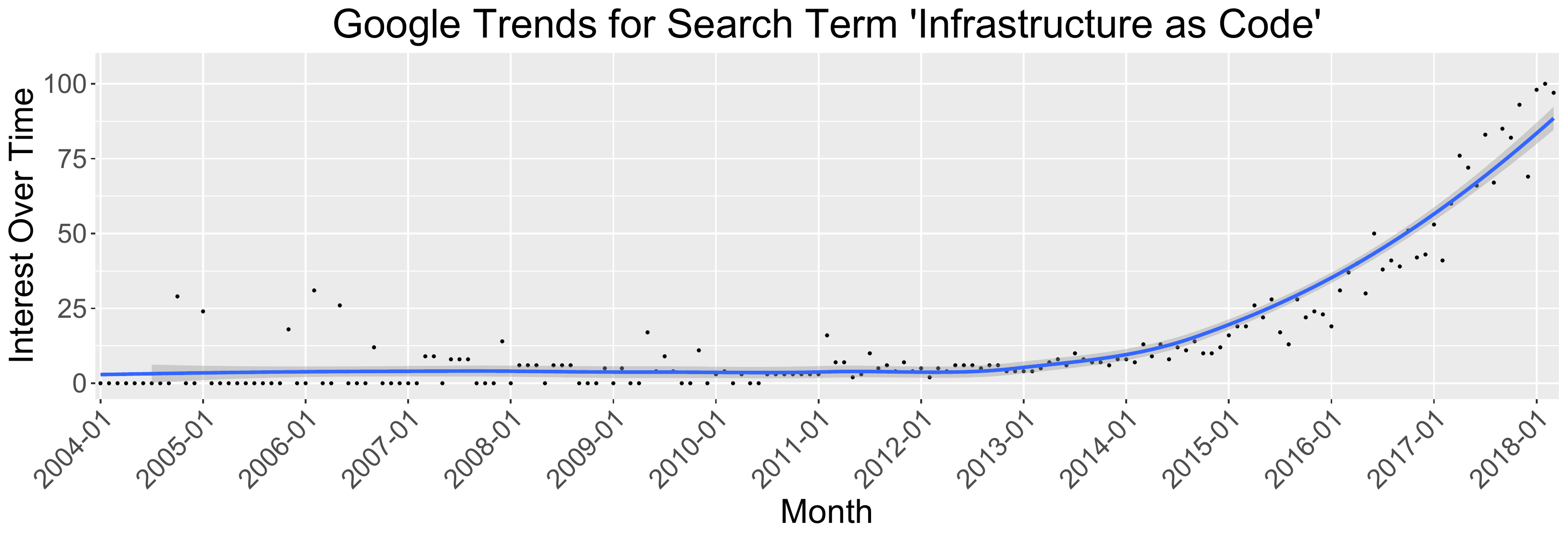}
\caption{Interest in IaC as a search topic since 2004 based on Google Trends data. Interest in IaC has steadily increased since 2015.}
\label{fig-intro-trend}
\end{figure}

Interest in the practice of IaC have grown amongst both: practitioners~\cite{cd:adage:parnin} and researchers~\cite{JiangAdamsMSR2015}~\cite{SharmaPuppet2016}. As shown in Figure~\ref{fig-intro-trend}, Google Trend~\footnote{https://trends.google.com/trends/explore?date=all\&q=Infrastructure\%20as\%20Code} data related to the search term `Infrastructure as Code', provides further evidence on how IaC as a topic has a growing interest. The x-axis presents months, and the y-axis presents the `Interest Over Time' metric determined by Google Trends. According to Figure~\ref{fig-intro-trend} interest in IaC has increased steadily after 2015. 

Even though interest in IaC is growing steadily, the current state of IaC research remains under-explored. A summary of existing literature in a particular research domain can help researchers to get an overview of the particular domain, and identify potential research topics that could benefit from systematic investigation. One strategy to summarize existing literature for a particular research domain is to conduct a systematic mapping study~\cite{sms:feldt}. Through a systematic mapping study, researchers can identify gaps, and can group existing research for a certain domain~\cite{sms:feldt}. The identified gaps can potentially direct future research in that particular domain~\cite{KITCHENHAM:IMP:SMS}. Researchers have conducted systematic mapping studies in numerous domains of software engineering, for example, in the domain of technical debt~\cite{debt:sms}, testing~\cite{testcode:sms}~\cite{test:effort:sms}, and software visualization~\cite{vis:val:sms}. Despite growing interest in IaC, we observe limited evidence of systematic mapping studies that have been conducted in the domain of IaC. We conduct a systematic mapping study in the domain of IaC that can be beneficial in two ways: (i) identify what research problems have already been addressed in the domain of IaC; and (ii) identify research problems that could benefit from further research.

\textit{The objective of this paper is to help researchers identify research areas related to infrastructure as code (IaC) by conducting a systematic mapping study of IaC-related research.}

We answer the following research questions: 

\begin{itemize}[leftmargin=*]
\item{\textbf{RQ1}: What topics have been studied in infrastructure as code (IaC)-related publications?}

\item{\textbf{RQ2}: What are the temporal publication trends for infrastructure as code (IaC)-related research topics?}

\item{\textbf{RQ3}: What are the temporal trends for the use of infrastructure as code (IaC)-related tools, as mentioned in IaC-related publications?}
\end{itemize}


We follow Petersen et al.~\cite{sms:feldt}'s guidelines, and conduct a systematic mapping study to identify which research topics are being studied in the domain of IaC. First, we search six scholar databases namely, IEEE Xplore~\footnote{http://ieeexplore.ieee.org/Xplore/home.jsp}, ACM Digital Library~\footnote{https://dl.acm.org/}, the IET Digital Library~\footnote{http://digital-library.theiet.org/}, Springer Link~\footnote{https://link.springer.com/}, ScienceDirect~\footnote{https://www.sciencedirect.com/}, and Wiley Online Library~\footnote{https://onlinelibrary.wiley.com/}. Using seven search strings, we obtain a set of 33,887 publications. By systematically applying inclusion and exclusion criteria~\cite{kitchen:guide:swe}, we obtain 31 IaC-related publications. We follow Kitchenham's guidelines~\cite{Kitchenham:Quality} to assess the quality of our set of 31 publications. We apply qualitative analysis~\cite{qual:coding:book} to generate topics from the content of the collected publications. Next, we investigate the overall and topic-wise temporal trends of the collected IaC-related publications. We also characterize the temporal trends of the use of IaC-related tools in our set of 31 publications.    

\textbf{Contributions}: We list our contributions as following:

\begin{itemize}
\item{A list of topics studied in IaC-related publications}; 
\item{An evaluation of the temporal trends for IaC-related publications}; and 
\item{An evaluation of the quality of IaC-related publications.}
\end{itemize}

We organize rest of the paper as following: in Section~\ref{bg-rel} we describe necessary background and related academic publications. We provide our methodology in Section~\ref{meth}. We provide our findings in Section~\ref{res}, and discuss possible implications of findings in Section~\ref{disc}. We list the limitations of our systematic mapping study in Section~\ref{threats}. Finally, we conclude our paper in Section~\ref{concl}.


\section{Background and Related Work}
\label{bg-rel}

In this section we first provide a brief background on IaC and systematic mapping studies. Then we describe related academic publications. 

\subsection{Background}
\label{bg}

In this section, we provide background on IaC and systematic mapping studies. 

\subsubsection{Background on Infrastructure as Code (IaC)}
\label{bg-iac}


Practitioners attribute the concept of infrastructure as code to Chad Fowler, in his blog published in 2013~\footnote{https://www.oreilly.com/ideas/an-introduction-to-immutable-infrastructure}. The phrase `as code' in IaC corresponds to applying traditional software engineering practices, such as code review and version control for IaC scripts~\cite{cd:adage:parnin}~\cite{Humble:2010:CD}. To automatically provision infrastructure, programmers follow specific syntax, and write configurations in a similar manner as software source code. IaC scripts use domain specific language (DSL)~\cite{ShambaughRehearsal2016}. Organizations that implement DevOps practices widely use commercial tools, such as Puppet, to implement IaC~\cite{Humble:2010:CD}~\cite{JiangAdamsMSR2015}~\cite{ShambaughRehearsal2016}. IaC scripts are also known as `configuration as code' scripts~\cite{SharmaPuppet2016}~\cite{Humble:2010:CD}.

We describe the typical work flow of IaC development as following: programmers make changes to the required IaC scripts and submit them to a version control system such as Git~\footnote{https://git-scm.com/}. Once changes are submitted, a build in the continuous integration (CI) tool, (e.g. Travis CI) is triggered. The CI tool runs the static analysis checks and test cases specified by the development team. If all the static analysis checks and tests pass, the CI tool integrates all the changes. 


\subsubsection{Background on Systematic Mapping Studies}
\label{bg-sms}

A systematic mapping study provides a `map' or an overview of a research area by (i) classifying papers and results based on relevant categories and (ii) counting the frequency of work in each of those categories. The output of a systematic mapping study is to identify the coverage of research studies in a particular area~\cite{sms:feldt}. Systematic mapping studies can be beneficial in identifying research studies relevant to that topics~\cite{sms:feldt}. Systematic mapping studies are different from systematic literature reviews (SLRs)~\cite{sms:feldt}, because unlike SLRs, systematic mapping studies are exploratory in nature, whereas, the purpose of SLRs is to provide a synthesized summaries to answer well-defined research questions~\cite{slr:sms:esem}. Systematic mapping studies have importance as these studies provide a basis for future research~\cite{KITCHENHAM:IMP:SMS}.

\subsection{Related Work}
\label{rel}

Our systematic mapping study is closely related to research studies on IaC, and prior research work that have conducted systematic mapping studies in other areas of software engineering. We briefly describe both in the following subsections: 

\subsubsection{Prior Research on IaC}
\label{rel-iac}

Our paper is related to empirical studies that have focused on IaC technologies, such as Puppet. Sharma et al.~\cite{SharmaPuppet2016} investigated anti-patterns in IaC scripts and proposed 13 implementation and 11 design anti-patterns. Hanappi et al.~\cite{Hanappi:2016:pupp:converge} investigated how convergence of Puppet scripts can be automatically tested and proposed an automated model-based test framework. Jiang and Adams~\cite{JiangAdamsMSR2015} investigated the co-evolution of IaC scripts and other software artifacts, such as build files and source code. They reported IaC scripts to experience frequent churn. Ikeshita et al.~\cite{Ikeshita:IaC:Reduction} proposed and evaluated a framework to reduce test suites for IaC. Weiss et al.~\cite{Weiss:Tortoise} proposed and evaluated `Tortoise', a tool that automatically corrects erroneous configurations in IaC scripts. Hummer at al.~\cite{Hummer:IaC} proposed a framework to enable automated testing of IaC scripts. 

We observe that researchers have a growing interest in the field of IaC. We take motivation from this observation, and conduct a systematic mapping study of IaC research in this paper. 


\subsubsection{Prior Research on Systematic Mapping Studies}
\label{rel-sms}

The use of systematic mapping studies is common in software engineering, for example in the domain of technical debt, domain specific languages, and software requirements. Li et al.~\cite{debt:sms} conducted a systematic mapping study with 94 publications related to technical debt management and observed the necessity of dedicated technical debt management tools in software engineering. Kosar et al.~\cite{dsl:sms} conducted a systematic mapping study with 390 publications related to domain specific languages (DSLs) and reported that the DSL community focuses more on the development of new techniques, instead of evaluating the effectiveness of the proposed DSL techniques. Novais et al.~\cite{evol:vis} studied 125 papers related to software evolution visualization and observed a lack of empirical research in the area of software evolution visualization. Jalali and Wohlin~\cite{global:sms} studied 77 papers related to the adoption of agile practices in global software engineering and reported that in majority of the papers agile practices were modified with respect to the context and situational requirements. Kitchenham~\cite{metrics:sms} studied 100 software metric-related publications and observed that empirical validation is a key focus of software metrics-related papers. Condori-Fernandez et al.~\cite{softevol:sms} reviewed 46 publications related to software requirement specification, and reported that understandability is the most commonly evaluated aspect of software requirement specification studies. Engstrom and Runeson~\cite{spl:sms} studied 64 publications on software product line testing, and advocated for stronger validation research methods to provide a better foundation for software product line testing. Paternoster et al.~\cite{startup:sms} extracted 213 software engineering practices from 43 publications related to software start-ups and reported that in software start-ups, software engineering work practices are chosen opportunistically, which are later adapted and configured. Elberzhager et al.~\cite{test:effort:sms}  studied 144 publications on reducing software testing efforts, and reported that researchers have focused more in the area of automation and prediction approaches. Yusifoglu et al.~\cite{testcode:sms} studied 60 publications on software test code engineering and observed that the two leading avenues of research in the area of software test code engineering are tools and methods. Seriai et al.~\cite{vis:val:sms} studied 87 publications related to validation of software visualization tools and observed the lack of maturity in validation of software visualization tools. Riaz et al.~\cite{maria:sms} studied 30 publication of software patterns and observed that software patterns in maintenance is the most commonly investigated domain in the research field of software patterns.  


The above-mentioned prior work illustrates the usage of systematic mapping studies in several areas of software engineering. We take motivation from these studies, and conduct a systematic mapping study in the area of IaC. Through our systematic mapping study we aim to identify the research areas that need attention in the field of IaC.


\section{Methodology}
\label{meth}

We conduct a systematic mapping study following the guidelines of Petersen et al.~\cite{sms:feldt}. In this section, we describe the methodology to conduct our systematic mapping study. The methodology is divided into four phases, which we describe in the following subsections:

\subsection{Phase One: Search}
\label{meth-search}

The first phase of finding IaC-related publications is to search the scholar databases. For our paper, we select six scholar databases following Kuhrmann et al.~\cite{emse:slr:guide}'s guidelines. These six scholar databases are: Institute of Electrical and Electronics Engineers (IEEE) Xplore~\footnote{http://ieeexplore.ieee.org/Xplore/home.jsp}, Association for Computing Machinery (ACM) Digital Library~\footnote{https://dl.acm.org/}, the Institution of Engineering and Technology (IET) Digital Library~\footnote{http://digital-library.theiet.org/}, Springer Link~\footnote{https://link.springer.com/}, ScienceDirect~\footnote{https://www.sciencedirect.com/}, and Wiley Online Library~\footnote{https://onlinelibrary.wiley.com/}. We select these six scholar databases as these databases are recommended for conducting systematic mapping studies and literature reviews~\cite{emse:slr:guide}. 

For searching the scholar databases, we construct a set of search strings. The construction process can be described as follows: 

\begin{itemize}
\item{Step-1: First, we perform an exploratory search in Google Scholar, using the string ``infrastructure as code''. We start with the string ``infrastructure as code'', as infrastructure as code (IaC) is the topic on which we conduct our systematic mapping study. Based on the search results, we observe that the string `infrastructure' can also refer to infrastructure in other disciplines such as civil engineering. Therefore, to limit our search scope in the area of IaC we added the string `software engineering', using which we derived the search string ``infrastructure as code AND software engineering''}. 
\item{Step-2: From the search results of Step-1, we observe that ``configuration as code'', is also used for ``infrastructure as code'' as a synonym~\cite{SharmaPuppet2016}. Similar to the search term ``infrastructure as code AND software engineering'', we also add the search string ``configuration as code AND software engineering''. IaC scripts are also referred to as configuration scripts~\cite{Humble:2010:CD}, so we created another search string ``configuration script AND software engineering''.} 
\item{Step-3: From the top five search results obtained from Step-1 and 2 we observe that publications that study IaC also use the keywords `devops', and `Puppet'. Therefore, as the third search string we use ``devops AND puppet''. As Ansible, CFEngine, and Chef are commonly used tools to implement IaC~\cite{SharmaPuppet2016}, we also include three more search strings: ``devops AND ansible'', ``devops AND chef'', and ``devops AND cfengine''. We do not consider `devops' as a search string, as this search string can yield search results that are applicable for DevOps only, such definitions and best practices of DevOps.} 
\end{itemize}
  
Altogether, we obtain the following seven search strings: 

\begin{itemize}[leftmargin=*]
\item{``infrastructure as code AND software engineering''}
\item{``configuration as code AND software engineering''}
\item{``configuration script AND software engineering''}
\item{``devops AND `puppet''}
\item{``devops AND `ansible''}
\item{``devops AND `chef''}
\item{``devops AND `cfengine''}
\end{itemize}       

We search each of the six scholar databases using the above-mentioned search strings. Our search process will result in a collection of publications that we filter using an inclusion and exclusion criteria, described in Section~\ref{meth-incl-excl}.


\paragraph{Quasi-Gold Set}\label{meth-quasi}: We use seven search strings in our search process. These search strings may yield search results that do not include IaC-related publications, which motivates us to validate the derived search strings. We validate our set of search strings by applying the `quasi-sensitivity' metric proposed by Zhang and Babar~\cite{Zhang:Babar:SLR}. The quasi-sensitivity (QS) approach validates if our set of search strings are sufficient to identify IaC-related publications. The QS metric requires a `quasi-gold' set of publications, which we identify as following: 

\begin{itemize}[leftmargin=*]
\item{First, we identify peer-reviewed publications that cite any of the following literature: `Continuous Delivery: Reliable Software Releases through Build, Test, and Deployment Automation'~\cite{Humble:2010:CD}, `Pro Puppet'~\cite{ProPuppet:Book}, `Infrastructure as Code: Managing Servers in the Cloud'~\cite{kief:iac:book}, and `DevOps for Developers'~\cite{devops:hatterman}. These publications discuss in details on how to implement the practices of DevOps and continuous deployment. As IaC is one of the fundamental pillars to implement continuous deployment and DevOps~\cite{Humble:2010:CD}, our assumption is that peer-reviewed publications that cite any of these books can be potentially relevant to conduct a systematic mapping study for IaC.}
\item{Second, we exclude publications that are not peer-reviewed, and not written in English.} 
\item{Third, we exclude publications that are not related to IaC by reading the titles of the collected publications. If we are unable to determine from the title, we read the publication completely. We use two raters to mitigate the subjectivity. The first and second author separately conducted this step. Upon completion, the agreements and Cohen's Kappa score~\cite{cohens:kappa} are recorded. The disagreements are resolved upon discussion. 

After completing this step we obtain a set of quasi-gold set of publications for our systematic mapping study.
}
\end{itemize}   

We calculate the quasi-sensitivity metric (QSM) using Equation~\ref{equ-meth-qs}. As a hypothetical example, if the total count of IaC-related publications in the quasi-gold set that is obtained using our search strings is 9, and the count of publications in our quasi-gold set is 10, then the quasi-sensitive score is 0.9.  


\begin{equation}
\text{QSM} = \\ \frac{\text{\# of publications from search strings, included in quasi-gold set}}{\text{\# of publications in quasi-gold set}}\label{equ-meth-qs}
\end{equation}

\subsection{Phase Two:  Inclusion and Exclusion Criteria}
\label{meth-incl-excl}

Search results obtained from using our search strings on the six databases contain irrelevant results that are out of scope for our research study. We filter those results using the following inclusion and exclusion criteria:

\begin{itemize}[leftmargin=*]
\item{Exclusion Criteria:
\begin{itemize}[leftmargin=*]
\item{Publications that are not peer-reviewed, for example, books}
\item{Publications are published before 2000. IaC-related concepts such as DevOps, continuous delivery, continuous deployment, and continuous integration are first introduced after 2000, and have gained in popularity since then. By selecting publications published on or after 2000, we assume to collect IaC-related publications needed for the systematic mapping study.}
\end{itemize}
}
\item{Inclusion Criteria:
\begin{itemize}[leftmargin=*]
\item{Publications must be written in English}
\item{Publications must be available for download}
\item{Title, Keywords, Abstract, and Introduction of the paper make it explicit that the paper is related to IaC}
\end{itemize}
}
\end{itemize} 

Upon applying the inclusion and exclusion criteria, we will obtain a set of publications that we use for our analysis. Before answering the RQs using our set of publications, we perform quality analysis to assess the quality of these publications, as described in Section~\ref{meth-qual}. 

\subsection{Phase Three:  Quality Analysis}
\label{meth-qual}

Kitchenham et al.~\cite{Kitchenham:Quality} proposed a set of criteria to evaluate the quality of software engineering publications. In their study, they used this criteria to assess if the quality of software engineering publications are increasing or decreasing as time progresses. A publication's higher quality score indicates that the publication of interest has stated their objectives clearly, has actionable findings, has discussed the limitations, and has clear presentation structure. We used Kitchenham et al.~\cite{Kitchenham:Quality}'s criteria set to assess the quality of our set of publications related to IaC:  

\begin{itemize}[leftmargin=*]
\item{Q1 (Aim): Do the authors clearly state the aim of the research?}
\item{Q2 (Units): Do the authors describe the sample and experimental units?}
\item{Q3 (Design): Do the authors describe the design of the experiment?}
\item{Q4 (Data Collection): Do the authors describe the data collection procedures and define the measures?}
\item{Q5 (Data Analysis): Do the authors define the data analysis procedures?}
\item{Q6 (Bias): Do the authors discuss potential experimenter bias?}
\item{Q7 (Limitations): Do the authors discuss the limitations of their study?}
\item{Q8 (Clarity): Do the authors state the findings clearly?}
\item{Q9 (Usefulness): Is there evidence that the Experiment/Quasi-Experiment can be used by other researchers/practitioners?}
\end{itemize} 

Based on the answers to each of the above-mentioned nine questions, a rater provides a score:  1 (not at all); 2 (somewhat); 3 (mostly); 4 (fully).  A higher score for each of this question, indicates that the authors of the paper have provided detailed descriptions, which can be helpful in replications and sound analysis~\cite{Kitchenham:Quality}~\cite{maria:sms}. As this process involves subjectivity, we use two raters who independently rated each question for each publication. We report the average score for each question and for each publication. 

Upon completion of this step, we obtain an assessment of quality for the collected publications that we use to answer our RQs. 

\paragraph{Threats Reported in IaC-related Publications}: When conducting research studies, validity threats may arise that either need to be accounted for or acknowledged as potential limitations. Explicit reporting of threats or limitations is indicative of high quality for an academic publication~\cite{Burcham:hotsos2017}~\cite{Kitchenham:Quality}. Furthermore, such investigation can guide future researchers to be aware of the what types of threats can occur if IaC-related research is conducted for certain topics. For each paper, we identify what types of threats have been reported using Wohlin et al.~\cite{wohlin:ese}'s four categories of validity threats:

\begin{itemize}[leftmargin=*]
\item{Conclusion Validity: Conclusion validity evaluates to which extent researchers have drawn conclusions from their analysis without violating statistical assumption and maintaining sufficient statistical power~\cite{wohlin:ese}.}
\item{Internal Validity: Internal validity evaluates to which extent researchers can make causal inferences from their empirical study~\cite{wohlin:ese}.}
\item{Construct Validity: Construct validity evaluates to which extent the experiment is measuring, what it is designed to measure~\cite{wohlin:ese}.}
\item{External Validity: External validity measures generalizability i.e. to which extent the reported results in the publication can be generalized in other contexts~\cite{wohlin:ese}.}
\end{itemize}

We do not make judgments about threats in the research that have not been reported by the authors.  

\subsection{Answer to Research Questions}
\label{meth-rq}

We describe the methodology to answer the three research questions as following: 

\subsubsection{Answer to RQ1: What topics have been studied in infrastructure as code (IaC)-related publications?}
\label{meth-rq1}

In RQ1 we focus on identifying the topics, which summarize the research avenues pursued in IaC-related publications. 


Answering RQ1 involves identifying topics that emerge from the IaC-related publications of interest. Each rater extracted sentences from the publication that convey important information about the topic of the publication (deemed ``raw text'').  Each rater applied qualitative analysis~\cite{qual:coding:book} to extract the topics of the sentences as verbatim phrases (deemed ``initial code''). These initial codes are abstracted to ``topics'' based upon commonalities observed in initial codes. 

\begin{figure*}
\includegraphics[width=0.95\textwidth]{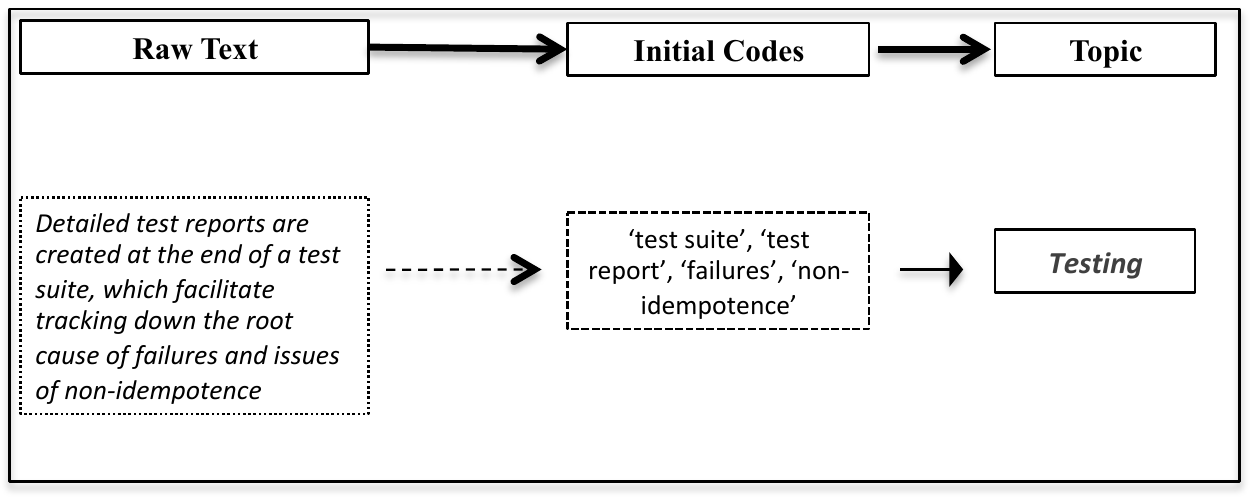}
\caption{An example of how we use qualitative coding to generate topics from the set of IaC-related publications.}
\label{figure-meth-code}
\end{figure*}

We use Figure~\ref{figure-meth-code} to illustrative our qualitative coding process. We first start with the extraction of raw text from a publication. Next, from the extracted `Raw Text' we derive initial codes. As demonstrated in Figure~\ref{figure-meth-code}, from the raw text `Detailed test reports are created at the end of a test suite, which facilitate tracking down the root cause of failures and issues of non-idempotence', we extract four initial codes: `test suite', `test report', `failures', and `non-idempotence'. Finally, we generate the topic `Testing' from the four initial codes. 


The process of generating topics is subjective, which we account for by deploying two raters. Two raters independently generate the topics from the collected publications. Two topic names that were determined to be synonyms were counted as an agreement. The disagreements are resolved upon discussion. Upon completion, we measure the agreement level on the generated topics, and the Cohen's Kappa score~\cite{cohens:kappa} is recorded.    

Answers to RQ1 will provide a list of topics that are studied in IaC-related publications. Each publication in our publication set can relate to more than one of the identified topics.  





\subsubsection{Answer to RQ2: What are the temporal publication trends for infrastructure as code (IaC)-related research topics?}
\label{meth-rq2}

We answer RQ2 using two approaches: first, we compute the overall trend of IaC-related publications by calculating how many publications are published in each year since 2000, related to IaC. Second, we compute the temporal trends exhibited for each identified topic. We accomplish this step by calculating the count of publications that belong to each topic are published each year. By using these two approaches we get two categories temporal of trends (i) an overall trend; and (ii) temporal trends of IaC-related publications per topic.  



\subsubsection{Answer to RQ3: What are the temporal trends for the use of infrastructure as code (IaC)-related tools, as mentioned in IaC-related publications?}
\label{meth-rq3}

The focus of RQ3 is to investigate what IaC tools are used to conduct IaC-related research. We take motivation from prior research that have conducted mapping studies to understand tool usage in other domains such as testing~\cite{sms:testing:tool} and global software engineering~\cite{sms:global:tool}. By answering RQ3 we can get insights on what types of tools have been reported in prior IaC-related publications and the corresponding task the tool have been used to accomplish. Let us consider the example of empirical studies. As a hypothetical example, let us assume that our analysis shows Puppet scripts to be used for conducting empirical analysis. Such information reveals the availability of Puppet scripts along with the repository sources (e.g. GitHub), and can be helpful for researchers who are interested in conducting empirical studies related to IaC. 


We answer RQ3 by analyzing each publication of our set. We determine a publication to use a tool $x$, if any of the following criteria is satisfied: 
\begin{itemize}[leftmargin=*]
\item{$x$ is used to implement a framework or methodology};
\item{$x$ is used to provision a system};
\item{scripts from $x$ is used to conduct an empirical study}; or
\item{scripts from $x$ is used to validate a proposed framework or methodology}
\end{itemize}   





\section{Results}
\label{res}

In this section, we first provide the count of publications that we derive using our search process, along with the publications that belong to our quasi-gold set. Next, we provide answers to the three RQs in the following sub-sections. 

In Table~\ref{table-res-search}, we report the count of publications for each scholar database. Altogether, we obtain 33,887 publications. We collect all these publications on December 30, 2017. From this set of 33,887 publications we remove duplicates, and separate out 14,015 publications. Next, we filter out publications that are written in English, and identify 10,567 publications. Finally, we remove 727 publications that are not peer-reviewed. After removing 727 publications we obtain our set of 9,840 publications. All the 9,840 publications are accessible and available for download.      

\paragraph{\textbf{Quasi-Gold Set}}: Altogether, we identify 11 IaC-related publications that belong to the quasi-gold set. The first author and second author respectively identify 9 and 11 publications nine of which were common between the two authors. Both authors agreed upon nine publications identified by the first author. The recorded Cohen's Kappa is 0.2. According to Landis and Koch~\cite{Landis:Koch:Kappa:Range}, the agreement level is `fair'. The first and second author resolve their disagreements by discussing their ratings and contents on the disagreed publications. Upon discussion between the first and second authors, two more publications are added to the set of nine publications identified by the first author. Using Equation~\ref{equ-meth-qs} reported in Section~\ref{meth-quasi}, we record a QS score of 1.0, for the collected publications. According to our QS score, using our set of search strings, we identify all publications in our quasi-gold set. The list of publications included in our quasi-gold set is available in Table 1 of Appendix. As an example, the publication `Cloud WorkBench: Benchmarking IaaS Providers based on Infrastructure-as-Code', is the first publication in our quasi-gold set and labeled as `QG1'.         



\begin{table}[]
\centering
\captionsetup{justification=centering}
\caption{Search Results for Scholar Databases}
\label{table-res-search}
\rowcolors{1}{}{lightgray}
\begin{tabular}{|p{4.0cm}|p{1.0cm}|}
\hline
\textbf{Scholar Database} & \textbf{Count} \\
\hline
\hline
ACM Digital Library &  420 \\
IEEE Xplore &  6,167 \\
ScienceDirect & 5,099 \\
Springer Link & 16,019 \\
Wiley Online & 3,793 \\
IET & 2,389 \\
\hline
\end{tabular}
\end{table} 

\begin{table*}[]
\centering
\captionsetup{justification=centering}
\caption{Quality Assessment of the 31 Publications}
\label{res-table-qual}
\rowcolors{1}{}{lightgray}
\tiny{
\begin{tabular}{|p{0.75cm}|p{1.0cm}|p{1.0cm}|p{1.0cm}|p{1.5cm}|p{1.5cm}|p{1.0cm}|p{1.5cm}|p{1.0cm}|p{1.5cm}|}
\hline
\textbf{Index} & \textbf{Q1} & \textbf{Q2} & \textbf{Q3} & \textbf{Q4}     & \textbf{Q5}   & \textbf{Q6} & \textbf{Q7} & \textbf{Q8} & \textbf{Q9}    \\
               & Aim         & Units     & Design      & Data Collection & Data Analysis & Bias        & Limitations & Clarity     & Usefulness    \\
\hline
\hline
S1  & \textbf{4.0} & \textbf{3.5} & \textbf{4.0} & \textbf{3.5} & \textbf{3.0} & \textbf{1.5} & \textbf{2.0} & 3.0 & \textbf{4.0} \\
S2  & \textbf{4.0} & \textbf{4.0} & \textbf{4.0} & \textbf{3.5} & \textbf{4.0} & \textbf{3.5} & \textbf{2.5} & \textbf{4.0} & \textbf{3.5} \\
S3  & 2.5 & 1.5 & 2.0 & 1.0 & 1.0 & 1.0 & 1.0 & 2.5 & \textbf{4.0} \\
S4  & 3.0 & \textbf{4.0} & \textbf{4.0} & \textbf{4.0} & \textbf{3.5} & \textbf{1.5} & \textbf{2.5} & \textbf{4.0} & \textbf{4.0} \\
S5  & \textbf{3.5} & \textbf{4.0} & \textbf{4.0} & \textbf{4.0} & \textbf{4.0} & \textbf{2.0} & \textbf{2.5} & \textbf{4.0} & \textbf{4.0} \\
S6  & 3.0 & \textbf{4.0} & \textbf{4.0} & \textbf{3.5} & \textbf{4.0} & \textbf{1.5} & \textbf{2.0} & \textbf{4.0} & \textbf{4.0} \\
S7  & 2.5 & 1.0 & 2.0 & 1.0 & 1.0 & 1.0 & 1.0 & 3.0 & \textbf{3.5} \\
S8  & \textbf{4.0} & \textbf{4.0} & \textbf{4.0} & \textbf{4.0} & \textbf{4.0} & \textbf{1.5} & \textbf{4.0} & \textbf{4.0} & \textbf{4.0} \\
S9  & \textbf{4.0} & \textbf{3.0} & \textbf{3.5} & 2.5 & \textbf{3.0} & \textbf{1.5} & \textbf{2.0} & 3.0 & \textbf{4.0} \\
S10 & \textbf{4.0} & \textbf{4.0} & \textbf{4.0} & \textbf{3.5} & \textbf{4.0} & \textbf{2.0} & \textbf{4.0} & \textbf{4.0} & \textbf{4.0} \\
S11 & 2.5 & 1.0 & 2.0 & 1.0 & 1.0 & 1.0 & 1.0 & 1.5 & 1.5 \\
S12 & 3.0 & \textbf{3.5} & \textbf{3.5} & \textbf{3.5} & \textbf{3.5} & 1.0 & 1.0 & 2.5 & 3.0 \\
S13 & 3.0 & 2.5 & 1.5 & 1.5 & 1.5 & 1.0 & 1.0 & \textbf{3.5} & \textbf{3.5} \\
S14 & 2.5 & 1.0 & 1.0 & 1.0 & 1.0 & 1.0 & 1.0 & 2.5 & 2.5 \\
S15 & 3.0 & 2.5 & 2.5 & 2.5 & 2.0 & 1.0 & 1.0 & 3.0 & 3.0 \\
S16 & 3.0 & 2.5 & 2.0 & 1.5 & 1.5 & 1.0 & 1.5 & 3.0 & 2.5 \\
S17 & 3.0 & 2.5 & 2.5 & 1.5 & 2.5 & 1.0 & 1.5 & 3.0 & 3.0 \\
S18 & 3.0 & \textbf{3.5} & \textbf{3.5} & \textbf{3.0} & \textbf{3.5} & 1.0 & \textbf{3.0} & \textbf{3.5} & \textbf{3.5} \\
S19 & 3.0 & 2.5 & \textbf{3.5} & \textbf{3.5} & \textbf{3.0} & \textbf{1.5} & \textbf{3.5} & \textbf{4.0} & 3.0 \\
S20 & 3.0 & 1.5 & \textbf{3.5} & 1.5 & 1.5 & 1.0 & 1.5 & \textbf{3.5} & 3.0 \\
S21 & 3.0 & \textbf{4.0} & \textbf{4.0} & \textbf{3.5} & \textbf{4.0} & \textbf{1.5} & 1.5 & \textbf{4.0} & \textbf{4.0} \\
S22 & 3.0 & 2.5 & 2.5 & 2.0 & 2.0 & \textbf{1.5} & \textbf{2.0} & 3.0 & 3.0 \\
S23 & 2.5 & \textbf{3.5} & 2.0 & \textbf{3.0} & \textbf{3.5} & \textbf{1.5} & 1.5 & \textbf{3.5} & 2.5 \\
S24 & 3.0 & \textbf{4.0} & \textbf{4.0} & \textbf{3.0} & \textbf{3.0} & \textbf{1.5} & \textbf{4.0} & \textbf{4.0} & \textbf{4.0} \\
S25 & 3.0 & \textbf{3.5} & \textbf{4.0} & \textbf{3.5} & \textbf{3.0} & 1.0 & 1.0 & \textbf{3.5} & \textbf{3.5} \\
S26 & 3.0 & \textbf{3.0} & \textbf{4.0} & 2.0 & 2.0 & \textbf{1.5} & 1.5 & 3.0 & 3.0 \\
S27 & \textbf{4.0} & 2.5 & 3.0 & 2.0 & 2.0 & 1.0 & 1.5 & \textbf{3.5} & 3.0 \\
S28 & 3.0 & 2.0 & 3.0 & 1.5 & 1.5 & 1.0 & 1.0 & 2.5 & 3.0 \\
S29 & 3.0 & 2.0 & 3.0 & 1.5 & 2.5 & 1.0 & \textbf{2.5} & 3.0 & 2.5 \\
S30 & 3.0 & 2.5 & \textbf{3.5} & 2.5 & \textbf{3.0} & 1.0 & 1.5 & \textbf{3.5} & 3.0 \\
S31 & 3.0 & 2.5 & \textbf{3.5} & 2.5 & 2.5 & 1.0 & 1.5 & 3.0 & 3.0 \\
\hline
\textbf{Avg.} & 3.1 & 2.8 & 3.1 & 2.5 & 2.6 & 1.3 & 1.9 & 3.3 & 3.3 \\
\hline
\end{tabular}
}
\end{table*}    

From the collected set of 9,840 publications, we determine if each of the publications are related to IaC. The first and second author individually complete this step to determine IaC-related publications. Both, the first and second author first read the titles of each of the 9,840 publications to determine if the publication related to IaC, identifying 85 and 98 publications, respectively. Next, by reading the abstract and the introduction of these publications, the first and second author respectively, identifies 36 and 39 publications, 26 of which were in common between the two authors.  We record a Cohen's Kappa of 0.81. According to Landis and Koch~\cite{Landis:Koch:Kappa:Range} the agreement level is `almost perfect'. The first and second author resolve their disagreements by discussing on their ratings and contents on the disagreed publications.   

After resolving disagreements between the first and second authors, we identify a set of 31 publications which we use to answer our three RQs. Each of the publications' names are listed in Table 2 of Appendix. We index each publications as `S\#', for example the index `S1' refers to the publication `Cloud WorkBench: Benchmarking IaaS Providers Based on Infrastructure-as-Code'. We acknowledge that this set of publications related to IaC is small. One possible reason can be attributed to the availability of artifacts: adoption of IaC is yet to become wide-spread to facilitate more research in the area of IaC. 
     
We also evaluate the publications' quality using the guidelines provided by Kitchenham et al.~\cite{Kitchenham:Quality}. We report our findings in Table~\ref{res-table-qual}. Each cell in the the Table corresponds to the average of the quality score determined by the two raters who are the first and second authors of the paper. For example, publication S1, has a quality score of 4.0 for the quality criteria Q1. Each quality criteria is followed by the theme of each quality criteria, as stated in Section~\ref{meth-qual}. For example, the quality criteria Q1 is related to the criterion of a publication's aim or goal being clearly stated. In Table~\ref{res-table-qual}, we report the average of scores for all 31 publications for each quality criteria in the `Avg.' row. The cells highlighted in bold indicate scores for a publication which has a score higher than the average for the quality criteria. For example, S1 has a higher score than that of the average score of all 31 publications for quality criteria Q1.     

For four quality checks, Q1, Q3, Q8, and Q9, the average score is higher than that of 3.0, which implies that our set of publications satisfy the checks of clearly stating aim of the publication; describing the design of the experiment; clearly stating findings; and identifying findings that are actionable for other researchers and practitioners. The average score is between 2.0 and 3.0 for quality checks Q2, Q4, and Q5. These three quality checks, respectively, presents the quality criterion of describing sample and experimental units; describing data collection procedures; and defining the data analysis procedures. Clear description of data collection and data analysis procedures can help in replicating research studies and in advancing the field of research in the area of IaC. Based on our findings, we recommend researchers who will conduct IaC-related research, to clearly define and describe their data collection and data analysis procedures.  

The scores are less than 2.0 for two quality checks Q6 and Q7 that, respectively, corresponds to discussion of potential experimental bias and to discussion of threats in the publication. Based on Kitchenham et al.'s guidelines~\cite{Kitchenham:Quality}~\cite{Kitchenham:Guideline:SWE}, research publications should clearly report potential experimenter bias, and the threats that are related to the research study. Future research studies can take our findings into account while conducting IaC-related research, and report the limitations and potential bias that may occur while conducting their research studies.   

In summary, our findings indicate that publications related to IaC can have actionable findings/suggestions for practitioners and researchers but lack necessary quality checks needed for proper and complete presentation of their findings. Based on our findings, we recommend researchers to report their IaC-related research findings by following the best practices suggested by Kitchenham et al.~\cite{Kitchenham:Quality}.   

\paragraph{Threats Reported in IaC-related Publications}: We also summarize which threats are reported in IaC-related publications. Altogether, we have considered four categories of threats: construct validity, conclusion validity, internal, and external validity. We observe that of the 31 publications, only 7 (22.5\%) explicitly report the publication's threat or limitations. A complete mapping between each publication and the reported threat categories for these seven publications is available in Table~\ref{rq4-res-table-pub}. In each cell we report if a category of threats is reported in a publication. For example, we observe that no Conclusion Validity was reported in S2.

Our findings suggest that IaC-related publications do not report the threats of their research studies adequately. We advocate for better reporting of research threats in IaC-related publications, following the guidelines of Wohlin et al.~\cite{wohlin:ese}.   


\begin{table*}[]
\centering
\captionsetup{justification=centering}
\caption{Reported Threats for Each Publication}
\label{rq4-res-table-pub}
\rowcolors{1}{}{lightgray}
\begin{tabular}{|p{1.5cm}|p{2.5cm}|p{3.0cm}|p{2.5cm}|p{2.5cm}|}
\hline
\textbf{Topic} & \textbf{Conclusion} & \textbf{Construction} & \textbf{Internal} & \textbf{External} \\
\hline
\hline
S2    & N & Y & Y & Y  \\
S5    & N & Y & N & Y  \\
S8    & N & Y & Y & N  \\
S9    & N & Y & N & N  \\
S10   & Y & Y & Y & Y  \\
S18   & N & Y & N & Y  \\
S24   & N & N & N & Y  \\
\hline
\end{tabular}
\end{table*} 


\subsection{Answer to RQ1: What topics have been studied in infrastructure as code (IaC)-related publications?}
\label{res-rq1}

We identify the topics that have been researched in the area of IaC by applying qualitative analysis. Through our qualitative analysis, we identify four topics. A publication can belong to multiple topics implying that the identified topics are not orthogonal to each other. The topics are: (1) Framework/Tool for infrastructure as code (Framework/Tool); (2) Use of infrastructure as code (Use of IaC); (3) Empirical study related to infrastructure as code (Empirical); and (4) Testing in infrastructure as code (Testing).   

A complete mapping between each of the 31 publication and their corresponding topic is available in Table~\ref{res-table-rq1-topic-paper}. We describe each topic, along with the count of publications for each  topic as following: 

\begin{table*}[]
\centering
\captionsetup{justification=centering}
\caption{Mapping Between Each Topic and Publication}
\label{res-table-rq1-topic-paper}
\rowcolors{1}{}{lightgray}
\begin{tabular}{|p{3.0cm}|p{10.0cm}|}
\hline
\textbf{Topic} & \textbf{Publication} \\
\hline
\hline
Framework/Tool    & S6, S10, S12, S13, S15, S18, S19, S20, S21, S22, S24, S25, S26, S29, S30, S31 \\
Use of IaC        & S1, S3, S7, S9, S14, S16, S17, S18, S23, S27, S28 \\
Empirical Study   & S2, S4, S5, S8, S10, S11, S23 \\
Testing           & S4, S5, S6, S11 \\
\hline
\end{tabular}
\end{table*} 

\begin{itemize}[leftmargin=*]
\item{\textbf{Framework/Tool for infrastructure as code (16)}: The most frequently studied topic in IaC-related publications is related to framework or tools. In these publications, authors propose a framework or a tool either to implement the practice of IaC or extend a functionality of IaC. We describe a few publications related to `Framework/Tool for IaC' briefly:  

Authors in S12 observed that a wide variety of reusable DevOps artifacts such as Chef cookbooks and Puppet modules are shared, but these artifacts are usually bound to specific tools. The authors proposed a novel framework that generates standard Topology and Orchestration Specification for Cloud Applications~\footnote{https://www.oasis-open.org/committees/tc\_home.php?wg\_abbrev=tosca}(TOSCA)-based DevOps artifacts to consolidate DevOps artifacts from different sources. Later, authors of S12 extend their work in S19, where they constructed a run-time framework using a open source tool-chain to support integration for a variety of DevOps artifacts. In S22 authors propose a the hidden master framework to assess the survivability of IaC scripts, when they are under attack. In S24 proposes a tool called ConfigValidator that validates IaC artifacts such as Docker images, by a writing rules to detect configurations. In S10, authors propose and evaluate Tortoise, which fixes configurations in Puppet scripts automatically. In S20, `Charon' is proposed to implement the practice of IaC. 

Existing tools can be limiting, which may be motivating researchers to propose framework or tools that mitigate these limitations. For example, in S20, the authors observed that existing commercial IaC tools make assumptions on configuration models, which may not suit the purpose of all IT organizations. Authors of publication S20 proposed Charon, a tool to implement IaC to mitigate this limitation.  
}

\item{\textbf{Use of infrastructure as code (11)}: Publications that relate to this topic discusses how IaC can be used in different domains of software engineering, such as monitoring of system and automated deployment of enterprise applications. We describe the publications that related to this topic briefly as following:  

S1 uses IaC to build a benchmark tool to assess the performance of cloud applications. Authors in S3 and S14 discusses how IaC can be used to implement DevOps. S7 focuses on how Ansible can be used to automatically provision an enterprise application. In S9, authors investigated the feasibility of using Puppet modules to deploy a software-as-a-service (SaaS) application. They observed that Puppet modules are adequate for provisioning SaaS applications, but comes with an extra layer of complexity. In S17 the authors propose `DevOpsLang' that uses Chef to automatically deploy a chat application. Authors in S18 proposes the ABS Modeling Language that uses IaC to deploy an e-commerce application. In S23, authors interview practitioners from 10 companies on the use of IaC in continuous deployment. The authors reported that IaC scripts have fundamentally changed how IT organizations are managing their servers using IaC. They also reported that similar to software code base IaC code bases churn frequently. Authors in S27 proposes Omnia that uses IaC to create a monitoring framework to monitor DevOps operations.

Our analysis suggests that use of IaC is not only limited to implement automated deployment and DevOps, but also to create monitoring applications for software systems. One possible explanation can be the ability to express system configurations in a programmatic manner using IaC scripts.  
}

\item{\textbf{Empirical study related to infrastructure as code (7)}: Publications that have conducted empirical studies related to IaC can be divided into two groups: publications focused on testing, and publications focused on non-testing issues. The three publications related to testing are S4, S5, and S11. The four publications that have conducted empirical analysis, but not are not focused on testing are S2, S8, S10, and S23. In S2, the authors observed IaC scripts churn frequently, making them susceptible to defects. Authors of S2 used 265 open source repositories to quantify the co-evolution of IaC scripts with software source code and software test code. In S8, authors studied code anti-aptterns that may cause maintainability issues for IaC script development and maintenance. Authors in S8 mined 4,621 open source Github repositories to identify code anti-patterns that can occur in IaC scripts. Authors of S10 proposed Tortoise, an automated program repair tool to fix configurations in Puppet scripts. In S23, the authors interviewed practitioners, and synthesized how practitioners from 10 companies use IaC scripts to implement continuous deployment. They observed IaC scripts to churn frequently, and are prone difficult to debug defects. 

}

\item{\textbf{Testing in infrastructure as code (4)}: We identify four publications that addresses the topic of testing for IaC scripts. From these four publications, we observe researchers' interest on testing the idempotence property of IaC scripts such as Chef and Puppet. In IaC it is expected that the deployed system converge into the desired state. Whether or not the deployed system has reached the desired state is called idempotence~\cite{Hummer:IaC}. In S6, the authors proposed a testing framework that test if Puppet scripts reach their convergence. S5 proposed a framework to test idempotence in IaC. Their approach used a state transition-based modeling approach to generate test cases to test idempotence for Chef scripts. In S4, the authors reported that the approach suggested in S5 generates too much test cases, and proposed an approach to reduce the amount of test cases to generate the test cases needed for testing of idempotence. The approach proposed in S4 combined testing and static verification approaches to generate test cases needed to test idempotence.  

Based on our analysis we observe the lack of empirical studies that focus on test coverage, test practices, and testing techniques. We advocate for research studies 
that can investigate other aspects of testing such as test coverage and testing practices.
}

\end{itemize}


\subsection{Answer to RQ2: What are the temporal publication trends for infrastructure as code (IaC)-related research topics?}
\label{res-rq2}

We answer RQ2 by first providing the count of publications that are published each year. We provide our findings in Table~\ref{res-table-rq2-1}. Even though our search process included publications starting from 2000, our earliest IaC-related publication, based on publication date is the year of 2012. The highest publication count is nine for the year 2017. 


\begin{table}[]
\centering
\captionsetup{justification=centering}
\caption{Frequency of IaC-related Publications}
\label{res-table-rq2-1}
\rowcolors{1}{}{lightgray}
\begin{tabular}{|p{2.0cm}|p{2.0cm}|}
\hline
\hline
\textbf{Year} & \textbf{Count}\\
\hline
2012 & 1\\
2013 & 3\\
2014 & 8\\
2015 & 4\\
2016 & 6\\
2017 & 9\\
\hline
\end{tabular}
\end{table}  

We also analyze the frequency of publications for each topic. We present our findings in Table~\ref{res-table-rq2-2}. We observe that for topics `Framework/Tool', `Use of IaC', and `Empirical Study', publication frequency increases after 2014, which is consistent with our overall trend in which we observe IaC-related publications to increase after 2014. We cannot make similar observations for other topics, as the count of publications may not be enough to report any existing trends.

\begin{table*}[]
\centering
\captionsetup{justification=centering}
\caption{Frequency of Publications per Year for Each Topic}
\label{res-table-rq2-2}
\rowcolors{1}{}{lightgray}
\begin{tabular}{|p{3.0cm}|p{1.0cm}|p{1.0cm}|p{1.0cm}|p{1.0cm}|p{1.0cm}|p{1.0cm}|p{1.0cm}|}
\hline
\textbf{Topic} & \textbf{2012} & \textbf{2013} & \textbf{2014} & \textbf{2015} & \textbf{2016} & \textbf{2017} \\
\hline
\hline
Framework/Tool   & 0 & 1 & 2 & 4 & 4 & 5 \\
Use of IaC       & 1 & 0 & 2 & 4 & 1 & 3 \\
Empirical Study  & 0 & 2 & 0 & 1 & 1 & 3 \\
Testing          & 0 & 2 & 0 & 0 & 1 & 1 \\
\hline
\end{tabular}
\end{table*}


\subsection{Answer to RQ3: What are the temporal trends for the use of infrastructure as code (IaC)-related tools, as mentioned in IaC-related publications?}
\label{res-rq3}

We answer RQ3 by reporting the IaC tools that are used to conduct the research reported in our collection of 31 publications. We first report the names of each IaC tool, and how many times each IaC tool was used in our set of publications in Table~\ref{res-table-rq3-1}. The `Tool' column presents the name of the tool, followed by a reference. The `Count Of Publications' column presents the count of publications that have used a certain IaC tool. We observe that 12 IaC tools were used in 31 publications, where the highest usage occurred for Chef: authors of 8 (25.8\%) IaC-related publications used Chef to conduct their research studies. 

\begin{table*}[]
\centering
\captionsetup{justification=centering}
\caption{Usage of IaC Tools}
\label{res-table-rq3-1}
\rowcolors{1}{}{lightgray}
\begin{tabular}{|p{6.0cm}|p{4.0cm}|}
\hline
\textbf{Tool} & \textbf{Count Of Publications}  \\
\hline
\hline
ABS Modeling Language~\cite{s18:abs}           & 1  \\
Ansible~\footnote{https://www.ansible.com/}    & 1  \\
Argon~\cite{s31:end2end}	                   & 2  \\
Charon~\cite{charon:s20}	                   & 1  \\
Chef~\footnote{https://www.chef.io/chef/}      & 8  \\
ConfigValidLang ~\cite{s24:configvalidator}    & 1  \\
DevOpsLang~\cite{s17:devopslang}               & 1  \\
Foreman~\footnote{https://www.theforeman.org/} & 1  \\
Juju~\footnote{https://jujucharms.com/}        & 3  \\
Omnia~\cite{s37:omnia}                         & 1  \\
Puppet~\footnote{https://puppet.com/}          & 6  \\
Vagrant~\footnote{https://www.vagrantup.com/}  & 1  \\
\hline
\end{tabular}
\end{table*}

We report the tool usage for publications included in each topic in Table~\ref{res-table-rq3-2}. Each tool is reported in the `Tool' column, and the count of each tool's usage in publications for each topic is represented in each cell. For topic `Framework/Tool' we observe a variety of tools to be used. In case of `Empirical Study' and `Testing' usage of tools are limited between Chef and Puppet. Our findings indicate that for conducting empirical studies in the area of IaC, scripts of popular tools such as Chef and Puppet may be more used than other tools such as Ansible or Juju. 

\begin{table*}[]
\centering
\captionsetup{justification=centering}
\caption{Usage of IaC Tools Amongst Topics}
\label{res-table-rq3-2}
\rowcolors{1}{}{lightgray}
\begin{tabular}{|p{3.0cm}|p{3.0cm}|p{2.0cm}|p{2.0cm}|p{1.5cm}|}
\hline
\textbf{Tool} & \textbf{Framework/Tool} & \textbf{Use of IaC} & \textbf{Empirical} & \textbf{Testing} \\
\hline
\hline
ABS Modeling Language            & 1 & 1 & 0 & 0 \\
Ansible    & 0 & 1 & 0 & 0 \\
Argon	                    & 2 & 0 & 0 & 0 \\
Charon                    & 1 & 0 & 0 & 0 \\
Chef      & 3 & 1 & 3 & 3 \\
ConfigValidLang      & 1 & 0 & 0 & 0 \\
DevOpsLang             & 0 & 1 & 0 & 0 \\
Foreman & 1 & 0 & 0 & 0 \\
Juju         & 3 & 0 & 0 & 0 \\
Omnia                       & 0 & 1 & 0 & 0 \\
Puppet          & 3 & 1 & 3 & 1 \\
Vagrant   & 1 & 1 & 0 & 0 \\
\hline
\end{tabular}
\end{table*}  

We also report the usage of IaC tools for year as reported in our set of 31 publications in Table~\ref{res-table-rq3-extra}. For the year 2012, we do not observe any publication in our set to use an IaC tool to conduct IaC-related research. Findings from Table~\ref{res-table-rq3-extra} suggest that From 2013 to 2017, use of two commercial tools Chef and Puppet, are higher than that of other IaC tools.     

\begin{table*}[]
\centering
\captionsetup{justification=centering}
\caption{Usage of IaC Tools per Year as Reported in Publications}
\label{res-table-rq3-extra}
\rowcolors{1}{}{lightgray}
\begin{tabular}{|p{3.0cm}|p{1.0cm}|p{1.0cm}|p{1.0cm}|p{1.0cm}|p{1.0cm}|p{1.0cm}|p{1.0cm}|}
\hline
\textbf{Tool} & \textbf{2012} & \textbf{2013} & \textbf{2014} & \textbf{2015} & \textbf{2016} & \textbf{2017} \\
\hline
\hline
ABS Modeling Language & 0 & 0 & 0 & 1 & 0 & 0 \\
Ansible               & 0 & 0 & 0 & 1 & 0 & 0 \\
Argon                 & 0 & 0 & 0 & 0 & 0 & 2 \\
Charon                & 0 & 1 & 0 & 0 & 0 & 0 \\
Chef                  & 0 & 2 & 2 & 3 & 0 & 1 \\
ConfigValidLang       & 0 & 0 & 0 & 0 & 0 & 1 \\
DevOpsLang            & 0 & 0 & 1 & 0 & 0 & 0 \\
Foreman               & 0 & 0 & 0 & 0 & 1 & 0 \\
Juju                  & 0 & 0 & 1 & 2 & 0 & 0 \\
Omnia                 & 0 & 0 & 0 & 0 & 0 & 1 \\
Puppet                & 0 & 0 & 0 & 2 & 3 & 1 \\
Vagrant               & 0 & 0 & 1 & 0 & 0 & 0 \\
\hline
\end{tabular}
\end{table*} 



\section{Discussion}
\label{disc}
In this section, we describe the implications of our systematic mapping study in the following sub-sections: 

\subsection{Research in IaC: State of the Art}

We have identified 31 IaC-related publications from 9,840 search results. Our findings indicate that as a research area, IaC is relatively new. Such observation however, is not surprising: in the field of software engineering, IaC has very recently getting popular with the increased popularity of DevOps and continuous deployment. As use of IaC gets popular in future, both in the open source and proprietary domain, we expect to see more research studies that will investigate different avenues of research for example, anti-patterns, barriers to adopt and use IaC, and code quality. 

We identify four topics with `Framework/Tool' being the most prevalent topic with respect to publication count. One possible explanation can be attributed to the usage of IaC in different teams. For example, finding existing IaC tools limiting, authors of S20 introduces a new IaC tool to implement the practice of IaC. Our conjecture is that depending on the needs of IT organizations, new frameworks or tools related to IaC are being proposed in publications. 

We also observe compared to other software engineering research areas, the frequency of publications related to empirical studies and testing are infrequent. We provide three possible explanations: 

\begin{itemize}[leftmargin=*]
\item{IT organizations have not adopted IaC at a wide scale and, as a result, empirical studies related to their experiences and challenges have not been reported}
\item{IT organizations that have adopted IaC are not open in sharing their experiences}
\item{Researchers do not have access to the necessary resources to perform empirical studies and other forms of research in the area of IaC}
\end{itemize} 

We do not observe any publication related to defects and security flaws. One possible explanation can be attributed to the lack of research resources. To conduct studies related to defects, validation, and verification for IaC researchers need access to relevant artifacts for example, scripts, bug reports, vulnerability reports etc., which may be non-trivial to obtain. 


\subsection{Variety of Tools}

In Table~\ref{res-table-rq3-2} we have reported 12 IaC tools that are used in our set of 31 publications. The three most frequent tools used are Chef, Puppet, and Juju. All these three tools are used for commercial purposes. Our findings suggest tools that are used commercially used for practitioners, such as Chef and Puppet, can be better-suited for future IaC-related research. Open source code repositories such as Github~\footnote{https://github.com/}, PuppetForge~\footnote{https://forge.puppet.com/} and Chef Cookbooks~\footnote{https://supermarket.chef.io/cookbooks}, can be a good source for conducting IaC research.       


\subsection{Potential Research Avenues in IaC}
Our findings reveal that researchers are yet to explore certain avenues for IaC. We do not observe publications to study defects and security flaws. We also observe lack of empirical studies conducted in the area of anti-patterns; only one publication studied anti-patterns in IaC scripts. We highlight potential research avenues that researchers may want to explore in the future: 

\begin{itemize}[leftmargin=*]
\item{\textbf{Anti-patterns}: Anti-patterns are recurring practices in software engineering that can have potential negative consequences~\cite{Brown:AP}. In our set of 31 publications, only one publication (S8) addressed the subject of anti-pattern. However, that study is limited to code anti-patterns. Researchers can explore what other anti-patterns can exist for IaC, for example, process anti-patterns, system architecture anti-patterns, security anti-patterns, and project management anti-patterns.}

\item{\textbf{Defect Analysis}: Defects in IaC scripts can have serious consequences, for example a defect in an IaC script caused a wide-scale outage for Gihtub~\footnote{https://github.com/blog/1759-dns-outage-post-mortem}. Based on our analysis, we do not observe existing IaC-related publications to study defects. We encourage researchers to investigate which characteristics of IaC correlate with defects, and how such defects can be mitigated.}

\item{\textbf{Security}: As IaC scripts are used to configure software systems and cloud instances at scale, an error that violates security objectives~\cite{nist:cia}, can compromise the entire system. In our set of 31 publications, we do not any publication that focus on security issues. Researchers can systematically study which security flaws are exhibited in IaC scripts, what are the consequences of such security flaws, and provide guidelines on how such flaws can be mitigated.}

\item{\textbf{Knowledge and Training}: Similar to any new technology, users of IaC, who are new to the technology can face challenges. What are the challenges in learning and implementing IaC, could be of interest to researchers. Such challenges can also provide recommendations on how course curriculum can be designed, so that students as well as practitioners are well-prepared for fulfilling IaC-related tasks in industry.}

\item{\textbf{Industry best practices}: Based on our analysis, we do not observe any research study that systematically characterizes the best practices for IaC implementation. Such characterization can be helpful for both: IT organizations that want to implement IaC, and for IT organizations who have already started implementing IaC. Synthesis of industry best practices exist for other domains such as, DevOps~\cite{rahman:csed:devsecops}, security~\cite{bsiim:2009}, and continuous deployment~\cite{cd:adage:parnin}~\cite{me:agile:cd}. Similar research initiatives to characterize industry best practices may also be beneficial for IaC adopters.}
\end{itemize}

\subsection{Towards Better Reporting of Research Findings}
As reported in Section~\ref{res}, none of the publication in our set has a perfect score of 4.0, for all quality checks. We also observe the majority of the publications to have actionable findings/suggestions for practitioners and researchers, but they not pass all quality checks. While reporting future IaC-related research results, researchers can take our findings into account. We advise researchers to follow guidelines provided by experts~\cite{Shaw:Good:SWE}~\cite{Kitchenham:Quality}~\cite{wohlin:ese}, when reporting their findings related to IaC research.  


\section{Threat to Validity}
\label{threats}

We discuss the limitations of our systematic mapping study as following: 

\begin{itemize}[leftmargin=*]

\item{\textbf{Internal Validity}: We acknowledge that our search process may not be comprehensive. As described in Section~\ref{meth}, we have used six scholar databases. We have not considered other scholar databases such as Scopus~\footnote{https://www.scopus.com/freelookup/form/author.uri}, which may include relevant IaC publications.  

Our use of seven search strings may also not be comprehensive, as the search strings may leave out IaC-related publications during our search process. We mitigated this threat by calculating the quasi-sensitivity metric (QSM), which yielded a score of 1.0.}

\item{\textbf{Conclusion Validity}: We apply a set of inclusion criteria to select which publications are related to IaC. We acknowledge that the process of selecting these publications can be subjective, with the potential of missing IaC-related publications. We mitigate the subjectivity by using two raters who individually determined which publications are related to IaC. 

We apply qualitative analysis to determine the topics that are being discussed in IaC-related publications. We determine these topics by extracting qualitative codes and following the guidelines of qualitative analysis~\cite{Stol:2016:GT:ICSE}. We acknowledge the process of generating topics can be subjective. We mitigate this limitation by using two qualitative raters.

}

\item{\textbf{External Validity}: Our analysis is dependent on our set of 31 publications collected on December 30, 2017. Furthermore, we have used certain scholar databases, which may not include all relevant publications for our paper. Due to the above-mentioned issues, generalizability of our findings can be limiting. We mitigate this threat by using six scholar databases recommended by Kurhamm et al.~\cite{emse:slr:guide} }

\end{itemize}


\section{Conclusion}
\label{concl}

IaC is a fundamental practice to implement continuous deployment. As adoption of DevOps amongst IT organizations gets increasingly popular, IaC can be an important research topic in the field of software engineering. A systematic mapping study can characterize existing research studies in the field of IaC and identify the open research areas in IaC. The goal of such study would be to help researchers in identifying potential research areas related to IaC. 

We accomplish this goal by conducting a systematic mapping study in the field of IaC. Using six scholar databases, we collect 31 publications related to IaC, which are systematically filtered from 33,887 publications. We generate four topics by performing qualitative analysis on the collected publications. These four topics are: (i) framework/tool for infrastructure as code; (ii) use of infrastructure as code; (iii) empirical study related to infrastructure as code; and (iv) testing in infrastructure as code. We observe the `Framework/Tool for infrastructure as code' to be the most prevalent topic, followed by `Use of infrastructure as code'. Our findings suggest that current research in IaC has mostly focused on implementing or extending the practice of IaC. We also observe 12 tools that are used in our set of 31 publications. The most frequently used tool is Chef, followed by Puppet.  

As defects and security flaws in IaC scripts can cause serious consequences, we advocate for research studies that addresses code quality issues such as defects and security flaws, along with exploring other research avenues. With respect to reporting research results, we advise researchers to follow the guidelines on writing good publications~\cite{Shaw:Good:SWE}~\cite{Kitchenham:Quality}, so that the expected quality checks of research studies are fulfilled. We hope our systematic mapping study will facilitate further research in the area of IaC.        



\section*{Supplementary material}

# 1. Appendix

Table 1: List of 11 Publications Included in the Quasi-Gold Set

| Index | Publication |
|-------|-------------|
| QG1 | Joel Scheuner, Jrgen Cito, Philipp Leitner, and Harald Gall. 2015. "Cloud WorkBench: Benchmarking IaaS Providers based on Infrastructure-as-Code". In Proceedings of the 24th International Conference on World Wide Web (WWW '15 Companion). ACM, New York, NY, USA, 239-242. |
| QG2 | Yujuan Jiang and Bram Adams. 2015. "Co-evolution of infrastructure and source code: an empirical study". In Proceedings of the 12th Working Conference on Mining Software Repositories (MSR '15). IEEE Press, Piscataway, NJ, USA, 45-55. |
| QG3 | M. Artac, T. Borovssak, E. Di Nitto, M. Guerriero and D. A. Tamburri, "DevOps: Introducing Infrastructure-as-Code," In Proceedings of the 2017 IEEE/ACM 39th International Conference on Software Engineering Companion (ICSE-C), Buenos Aires, 2017. |
| QG4 | Hummer, W., Rosenberg, F., Oliveira, F., Eilam, T., "Testing Idempotence for Infrastructure as Code", IIn Proceedings of the International Conference on Middleware (Middleware'13), 2013. |
| QG5 | Oliver H., Waldemar H., and Schahram D., "Asserting reliable convergence for configuration management scripts", SIGPLAN Not. 51, 10 (October 2016). |
| QG6 | T. Sharma, M. Fragkoulis, and D. Spinellis, "Does your configuration code smell?", In Proceedings of the 13th International Conference on Mining Software Repositories (MSR '16). ACM, New York, NY, USA. |
| QG7 | J. Hintsch, C. Grling and K. Turowski, "Modularization of Software as a Service Products: A Case Study of the Configuration Management Tool Puppet," In Proceedings of the 2015 International Conference on Enterprise Systems (ES), Basel, 2015. |
| QG8 | Waldemar Hummer, Florian Rosenberg, Fabio Oliveira, and Tamar Eilam. 2013. "Automated testing of chef automation scripts", In Proceedings Demo & Poster Track of ACM/IFIP/USENIX International Middleware Conference (MiddlewareDPT '13). ACM, New York, NY, USA. |
| | Continued on next page |

| Index | Publication |
|-------|-------------|
| QG9 | B. Adams and S. McIntosh, "Modern Release Engineering in a Nutshell – Why Researchers Should Care," In Proceedings of the 2016 IEEE 23rd International Conference on Software Analysis, Evolution, and Reengineering (SANER), Suita, 2016. |
| QG10 | D. Spinellis, "Don't Install Software by Hand," in IEEE Software, vol. 29, no. 4, pp. 86-87, July-Aug. 2012. |
| QG11 | J. Hintsch, C. Grling and K. Turowski, "Modularization of Software as a Service Products: A Case Study of the Configuration Management Tool Puppet," In Proceedings of the 2015 International Conference on Enterprise Systems (ES), Basel, 2015. |

Table 2: List of 31 Publications for Systematic Mapping Study

| Index | Publication |
|-------|-------------|
| S1 | Joel Scheuner, Jrgen Cito, Philipp Leitner, and Harald Gall. 2015. "Cloud WorkBench: Benchmarking IaaS Providers based on Infrastructure-as-Code". In Proceedings of the 24th International Conference on World Wide Web (WWW '15 Companion). ACM, New York, NY, USA, 239-242. |
| S2 | Yujuan Jiang and Bram Adams. 2015. "Co-evolution of infrastructure and source code: an empirical study". In Proceedings of the 12th Working Conference on Mining Software Repositories (MSR '15). IEEE Press, Piscataway, NJ, USA, 45-55. |
| S3 | M. Artac, T. Borovssak, E. Di Nitto, M. Guerriero and D. A. Tamburri, "DevOps: Introducing Infrastructure-as-Code," In Proceedings of the 2017 IEEE/ACM 39th International Conference on Software Engineering Companion (ICSE-C), Buenos Aires, 2017. |
| S4 | Ikeshita, K., Ishikawa, F., Honiden, S., Gabmeyer, S., Johnsen, E., "Test Suite Reduction in Idempotence Testing of Infrastructure as Code", In Proceedings of the International Conference on Tests and Proofs (TAP'17), 2017. |

**Table 2 – continued from previous page**

| Index | Publication |
|---|---|
| S5 | Hummer, W., Rosenberg, F., Oliveira, F., Eilam, T., "Testing Idempotence for Infrastructure as Code", IIn Proceedings of the International Conference on Middleware (Middleware'13), 2013. |
| S6 | Oliver H., Waldemar H., and Schahram D., "Asserting reliable convergence for configuration management scripts", SIGPLAN Not. 51, 10 (October 2016). |
| S7 | Nishant Kumar Singh, S. Thakur, H. Chaurasiya and H. Nagdev, "Automated provisioning of application in IAAS cloud using Ansible configuration management," in Proceedings of the 2015 1st International Conference on Next Generation Computing Technologies (NGCT), Dehradun, 2015. |
| S8 | T. Sharma, M. Fragkoulis, and D. Spinellis, "Does your configuration code smell?", In Proceedings of the 13th International Conference on Mining Software Repositories (MSR '16). ACM, New York, NY, USA. |
| S9 | J. Hintsch, C. Grling and K. Turowski, "Modularization of Software as a Service Products: A Case Study of the Configuration Management Tool Puppet," In Proceedings of the 2015 International Conference on Enterprise Systems (ES), Basel, 2015. |
| S10 | Aaron Weiss, Arjun Guha, and Yuriy Brun, "Tortoise: interactive system configuration repair", In Proceedings of the 32nd IEEE/ACM International Conference on Automated Software Engineering (ASE 2017). IEEE Press, Piscataway, NJ, USA. |
| S11 | Waldemar Hummer, Florian Rosenberg, Fabio Oliveira, and Tamar Eilam. 2013. "Automated testing of chef automation scripts", In Proceedings Demo & Poster Track of ACM/IFIP/USENIX International Middleware Conference (MiddlewareDPT '13). ACM, New York, NY, USA. |
| S12 | Johannes Wettinger, Uwe Breitenbcher, and Frank Leymann. "Standards-Based DevOps Automation and Integration Using TOSCA", In Proceedings of the 2014 IEEE/ACM 7th International Conference on Utility and Cloud Computing (UCC '14). IEEE Computer Society, Washington, DC, USA. |
| | |

**Table 2 – continued from previous page**

| Index | Publication |
|-------|-------------|
| S13 | B. Adams and S. McIntosh, "Modern Release Engineering in a Nutshell – Why Researchers Should Care," In Proceedings of the 2016 IEEE 23rd International Conference on Software Analysis, Evolution, and Reengineering (SANER), Suita, 2016. |
| S14 | D. Spinellis, "Don't Install Software by Hand," in IEEE Software, vol. 29, no. 4, pp. 86-87, July-Aug. 2012. |
| S15 | M. Miglierina, "Application Deployment and Management in the Cloud," In Proceedings of the 2014 16th International Symposium on Symbolic and Numeric Algorithms for Scientific Computing, Timisoara, 2014. |
| S16 | Kecskemeti G., Kertesz A., Marosi A.C., "Towards a Methodology to Form Microservices from Monolithic Ones", In Proceedings of the Euro-Par 2016: Parallel Processing Workshops. Euro-Par 2016, 2016. |
| S17 | Wettinger, J., Breitenbcher, U., Leymann, F., "DevOpSlang  Bridging the Gap between Development and Operations", In Service-Oriented and Cloud Computing, 2014. |
| S18 | de Gouw, S., Lienhardt, M., Mauro, J., Nobakht, B., Zavattaro, G., "On the Integration of Automatic Deployment into the ABS Modeling Language", In Service Oriented and Cloud Computing, 2015. |
| S19 | Johannes W., Uwe B., Oliver K., and Leymann F., "Streamlining DevOps automation for Cloud applications using TOSCA as standardized metamodel", In Future Generation Computer Systems 56, C (March 2016). |
| S20 | Eelco Dolstra, Rob Vermaas, and Shea Levy, "Charon: declarative provisioning and deployment", In Proceedings of the 1st International Workshop on Release Engineering (RELENG '13). IEEE Press, Piscataway, NJ, USA. |
| S21 | P. Kathiravelu and L. Veiga, "SENDIM for Incremental Development of Cloud Networks: Simulation, Emulation and Deployment Integration Middleware," In Proceedings of the 2016 IEEE International Conference on Cloud Engineering (IC2E), Berlin, 2016. |

**Table 2 – continued from previous page**

| Index | Publication |
|-------|-------------|
| S22 | T. Karvinen and S. Li, "Investigating survivability of configuration management tools in unreliable and hostile networks," In Proceedings of the 2017 3rd International Conference on Information Management (ICIM), Chengdu, 2017. |
| S23 | Chris Parnin, Eric Helms, Chris Atlee, Harley Boughton, Mark Ghattas, Andy Glover, James Holman, John Micco, Brendan Murphy, Tony Savor, Michael Stumm, Shari Whitaker, and Laurie Williams, "The Top 10 Adages in Continuous Deployment", In IEEE Softw. 34, 3 (May 2017). |
| S24 | Salman Baset, Sahil Suneja, Nilton Bila, Ozan Tuncer, and Canturk Isci, "Usable declarative configuration specification and validation for applications, systems, and cloud", In Proceedings of the 18th ACM/IFIP/USENIX Middleware Conference: Industrial Track (Middleware '17). |
| S25 | Lwakatare, L., Kuvaja, P., Leymann, F., Oivo, M., "Dimensions of DevOps", In Proceedings of the Agile Processes in Software Engineering and Extreme Programming, 2015. |
| S26 | J. Wettinger, V. Andrikopoulos and F. Leymann, "Automated Capturing and Systematic Usage of DevOps Knowledge for Cloud Applications," In Proceedings of the 2015 IEEE International Conference on Cloud Engineering, Tempe, AZ, 2015. |
| S27 | Marco Miglierina and Damian A. Tamburri, "Towards Omnia: A Monitoring Factory for Quality-Aware DevOps", In Proceedings of the 8th ACM/SPEC on International Conference on Performance Engineering Companion (ICPE '17 Companion). ACM, New York, NY, USA. |
| S28 | M. Virmani, "Understanding DevOps & bridging the gap from continuous integration to continuous delivery," In Proceedings of the Fifth International Conference on the Innovative Computing Technology (INTECH 2015), Galcia, 2015. |
| S29 | Aiftimiei C., Costantini A., Bucchi R., Italiano A., Michelotto D., Panella M., Pergolesi M., Saletta M., Traldi S., Vistoli C., Zizzi G., Salomoni D., "Cloud Environment Automation: from infrastructure deployment to application monitoring", In Journal of Physics: Conference Series, 2017. |

**Table 2 – continued from previous page**

| Index | Publication |
|-------|-------------|
| S30 | J. Sandobalin, E. Insfran and S. Abrahao, "An Infrastructure Modelling Tool for Cloud Provisioning," In Proceedings of the 2017 IEEE International Conference on Services Computing (SCC), Honolulu, HI, 2017. |
| S31 | Sandobalin, J., Insfran, E., Abrahao, S., "End-to-End Automation in Cloud Infrastructure Provisioning", In Proceedings of the Information Systems Development: Advances in Methods, Tools and Management Conference (ISD2017), Larnaca, Cyprus, 2017. |



\section{References}
\bibliographystyle{elsarticle-num} 
\bibliography{ist}

\end{document}